\documentclass[preprint,amsmath,amssymb,prl]{revtex4-1}
\usepackage{graphicx,color}

\begin{document}

\title{Graphene-based synthetic antiferromagnets and ferrimagnets}

\author{P. Gargiani$^{1}$}
\author{R. Cuadrado$^{2}$}
\author{H. B. Vasili$^{1}$}
\author{M. Pruneda$^{2}$} 
\author{M. Valvidares$^{1}$}

\affiliation{$^{1}$ ALBA Synchrotron Light Source, E-08290 Cerdanyola del Valles, Barcelona, Spain}
\affiliation{$^{2}$ ICN2 CSIC, Institut Catala de Nanociencia i Nanotecnologia, Campus UAB, 08193 Bellaterra, Barcelona, Spain}

\begin{abstract}
Hybrid Graphene/magnetic structures offer a unique playground for fundamental research, and opportunities for emerging technologies. Graphene-spaced ultrathin structures with antiferromagnetic exchange-coupling (AFC) seem a relevant scenario, analogous to that of conventional metallic-multilayer devices.
Unfortunately, the AFC found so far between bulk magnetic single crystals and Graphene-spaced adatoms, clusters or molecules either requires low temperatures, is too weak, or of complex nature, for realistic exploitation.
Here we show theoretically and experimentally that a strong perpendicular AFC can be established in ultrahin-film structures such as Fe/Gr/Co on Ir(111), first-time enabling {\it Graphene-based synthetic antiferromagnet and ferrimagnet} materials with unprecedented magnetic properties and appearing suitable for applications.
Remarkably, the established AFC is robust on structure thicknesses, thermally stable up to room temperature, very strong but field-controllable, and occurs in perpendicular orientation with opposite high remanent layer magnetizations.
Our atomistic first-principle simulations provide further ground for the feasibility of Graphene-mediated AFC ultra-thin film structures, revealing that Graphene acts not only as mere spacer but has a direct role in sustaining antiferromagnetic superexchange-coupling between the magnetic layers.
These results provide a path for the design of unique and ultimately-thin synthetic antiferromagnetic structures, which seem exciting for fundamental nanoscience studies or for potential use in Graphene-spintronics applications.
\end{abstract}
\maketitle

Hybrid Graphene/magnetic structures display a variety of physical phenomena and properties such as room-temperature long-spin lifetimes, spin filtering and tunnel magneto-resistance \cite{Karpan2007,Li2014a,Park2014,Martin2015}, which could yield a range of innovative {\it graphene spintronic} technologies \cite{Novoselov2012,Roche2015}. Also remarkable, occurrence of antiferromagnetic (AF) Graphene-mediated exchange coupling was observed between bulk magnetic single crystals and a Graphene-spaced molecular layer incorporating magnetic ligand atoms \cite{Hermanns2013,Garnica2013}. AFC between magnetic bulk single crystals and Graphene-spaced magnetic adatoms has been more recently studied, and evidenced to dilute or evolve onto a complex coupling when increasing size from adatoms to small clusters \cite{Barla2016}. Assessing the possibility to realize exchange coupled magnetic thin-films across a single graphene layer appears of primary importance towards the realization of spintronic devices based on graphene. This is particularly true for perpendicular magnetization geometry which appears technologically better suited for smaller, higher-stability, magnetic devices. However, the Graphene-mediated exchange coupling playground remains largely unexplored, and unexploited, largely due to the complex fabrication process, and the limitations set by epitaxy constraints and the high temperatures needed for in-situ growth of CVD Graphene.

Moving away from bulk magnetic materials into thin film structures or nanostructures, is key to achieve the development of novel Graphene-based AFC systems with robust properties, as demonstrated hereafter. By using only magnetic ultrathin films, one can achieve strong magnetic anisotropies and exchange coupling (or interesting Dzyaloshinskii--Moriya interactions) via interfacial engineering \cite{Johnson:RPP:1996,Fert2016}, whereas including a bulk (single crystal) magnetic layer dilutes the effect of interfacial contributions. One suitable path to shift away from bulk magnetic materials and pursue investigations on such virgin ground is the intercalation approach widely used in the 70s-80s \cite{Dresselhaus1981}, and more recently employed to intercalate magnetic layers below Graphene \cite{Wiesendanger2011,Rougemaille2012}. This approach is followed here for the fabrication of single-layer Graphene spaced ultrathin FM1/Gr/FM2 structures to explore Graphene mediated exchange coupling and to pursue the realization of robust Graphene-based AFC thin-film heterostructures suitable for applications.
It is worth recalling that synthetic (also called artificial) ferrimagnetic and antiferromagnetic materials based on metallic multilayers were developed in the late 80s, with the discovery of exchange coupling in multilayers \cite{Grunberg1986} and relevant associated phenomenology such as giant-magneto-resistance \cite{Fert1988,Grunberg1989} and oscillatory behavior \cite{Parkin1990}. More recently, antiferromagnetic materials, including synthetic ferrimagnetic and antiferromagnetic structures, are receiving renewed attention because their potential use in magnetic information storage enabling faster racetrack memories \cite{Parkin2015}, as all-optical materials \cite{Mangin2014}, as Pt free TMJ memories \cite{Dieny2016}, or even facilitating an AF-based spintronic technology \cite{Park2011,Jungwirth2016}.

\begin{figure*}
\includegraphics[width=0.95\textwidth]{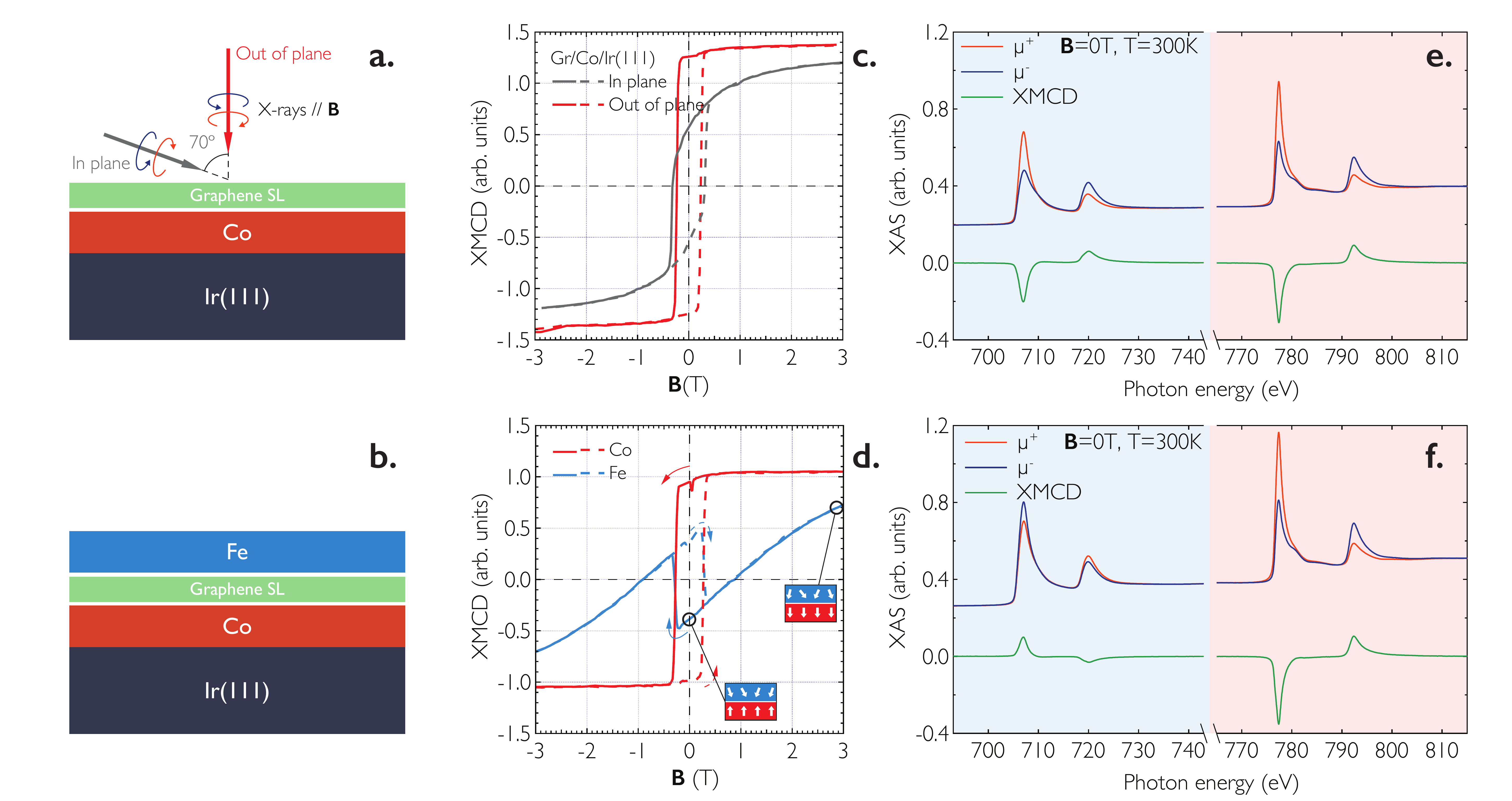}
\centering
\caption{\label{fig:co_loop}\textbf{System properties as deduced by element-sensitive hysteresis loop and XMCD.} Experimental geometry (a) for the in-plane and out-of-plane hysteresis measurements and a cartoon (b) depicting the Fe/Gr(SL)/Co/Ir(111) multilayer sample. Element-specific hysteresis loop measured along the Co film out-of-plane easy magnetization axis for a (c) Gr/Co[1.9 ML]/Ir(111) and (d) the same sample covered with Fe[1.6 ML] measured at T=300K as the L$_3$ XMCD intensity maximum normalized by the pre-edge value as a function of the applied field. The AF-ordering of the Fe with respect to the PMA Co layer is clearly denoted by the magnetization-loop sign inversion at low field, as depicted by the cartoon in the figure inset. XMCD spectra measured on the Fe[1.6 ML]/Gr/Co[1.9 ML]/Ir(111) at f) H=6T and g) H=0T after having ramped the magnetic field to H=+6T. The Fe/Co AFC is evident as the XMCD sign inverts between the two atomic species.}
\end{figure*}

A magnetic-Graphene multilayer composed by an intercalated Gr/Co/Ir(111) ultrathin structure coupled to a Fe overlayer displays a remarkable AFC at room temperature, as revealed by the set of element-sensitive X-ray Magnetic Circular Dichroism (XMCD) measurements displayed in Figure \ref{fig:co_loop}. As deduced from room-temperature field-dependent XMCD intensities reported in Fig.\ref{fig:co_loop}(c), the Gr/Co[1.9ML]/Ir(111) intercalated film presents high perpendicular magnetic anisotropy (PMA) and a large coercive field of $H_{c}$=0.27$~$T, together with a 92\% magnetic remanence along the out-of-plane easy-axis. After subsequent evaporation of the Fe overlayer, the resulting Fe[1.6ML]/Gr/Co[1.9ML]/Ir(111) multilayer exhibits a room-temperature field-controlled AFC which is clearly evidenced by the Fe and Co element-specific hysteresis loops reported in Fig.\ref{fig:co_loop}(d): as the applied magnetic field is decreased from the positive value of +3T to zero the Co magnetization (red continuos line) exhibits a high remanent state, whereas the Fe magnetization (blue continuos line) is progressively reduced and eventually crosses zero at B$\simeq{}0.8\,$T, reversing its sign at lower fields yielding an antiparallel alignment at H=0, i.e. AFC of the Fe and Co out-of-plane magnetizations. The remanent AF state of the system is also highlighted by the XMCD spectrum in Fig.\ref{fig:co_loop}(f), again denoting an opposed alignment of the Fe and Co magnetizations. For a negative applied field attaining the Co layer coercive field, the Co layer magnetization switches, driving the Fe layer magnetization on a corresponding abrupt jump and preserving the AF alignment.
Under increased reverse applied field, the Fe perpendicular magnetization decreases monotonically down to zero. This defines a critical applied field which destroys the AF alignment, hence compensating the AF exchange coupling energy between Fe and Co perpendicular magnetizations. This compensation field is indicative of the strength of the AFC, and for the Fe[1.6ML]/Gr/Co[1.9ML]/Ir(111) sample amounts to H=0.9$\,$T at T=300K. For larger negative field, the Fe magnetization gets progressively re-oriented along the field direction on a rather linear manner. Even at the maximum available external field of 6$\,$T, the magnetic saturation of the Fe layer is not completely reached, as evidenced by the non-horizontal slope of the Fe magnetization curve. Still, we roughly estimate that the remanent state shows a 60\% or higher polarization of the Fe layer.
One likely hypothesis would be that this behavior is related to considerable 3D-film morphology, giving place to a distribution of switching fields and/or weekly coupled magnetic regions or grains, but other possibilities should not be excluded taking into account that the magnetization reversal of AF exchange coupled systems has been demonstrated to yield complex phenomenologies \cite{Dieny1990}.

It is worth discussing in detail on how the Fe/Gr/Co/Ir(111) multilayer have been fabricated, as this provides a path that seem of general applicability and relevance in views of fabricating other FM/Gr hetero-structures with tailored magnetic properties. The growth of the Co film below the graphene layer was realized by the thermally-activated intercalation of Co adatoms e-beam deposited on top of a single high-quality layer of Gr/Ir(111) grown in-situ by CVD (see methods and supp. material S2). In order to achieve high PMA, remnant magnetization above 80\% and high coercivity, the thickness of the Co layer deposited was optimized to about 1.9$\pm$0.2 monolayer (see supp. material S4). The cobalt intercalation below the Graphene carpet is activated at temperature exceding 500K, and the optimal PMA/high magnetic remanence of the Co layer is obtained at T=700K. The intercalation process has been previously shown to take place at Graphene point defects and wrinkles \cite{Vlaic2014}, and remarkably it does not affect the Graphene layer quality enabling the fabrication of a high-quality graphene-protected Co/Ir(111) layer (see supp. material S7) \cite{Vlaic2014}. At a structural level, the sharp and low background Low Energy Electro-Diffraction patterns (fig S3) are indicative of a lattice matched Co growth on Ir(111). 

In what respect the resulting Fe overlayer morphology in our systems, we can infer some valuable information from the LEED patterns registered, which do not offer the capability of local probes to image with atomic resolution but in exchange offer a unique statistically averaged information over a large area \cite{henzler:SS:1996}. The Fe/Gr/Co/Ir(111) LEED patterns analysis (see supp. material S2) provides evidence for Fe islands having average lateral dimensions of the order of 6-8 nm for a 1~ML Fe film, suggesting that a large surface coverage can be attained already at few MLs coverage.
These results seem to agree with what might be expected from our calculated absorption energies (see below), and experimental evidences obtained by STM on similar surfaces. Firstly, the growth of Fe on Gr/Ir111 beyond an initial 3D growth phase, results on high-island density and high surface coverage  (55$\%$ at 2ML, which extrapolates to 80-90$\%$ for 3MLs) due to long range repulsive interactions \cite{Binz2012}; 
additionally, the computed adsorption energies of Fe on Gr/Co/Ir (see below) being comparable to the Fe on Fe ones, should favor even more the surface coverage over the incorporation of deposited atoms atop of the islands.

Summarizing the interpretation of our LEED data together with these considerations, it seems reasonable to expect that our Fe films with equivalent thickness ranging between 2 to 4 MLs have a 3D thin film growth with a full or almost full surface coverage coexisting with a certain amount of 3D multilevel islands. Furthermore, it is interesting to mention that the unique growth morphology of Fe (also Co) layers on Graphene enables its control by deposition temperature and deposited thickness. Finally, atomically flat magnetic overlayers on Graphene have been demonstrated by pulsed laser assisted growth \cite{Vo-Van2010}, which might be ascribed to a larger instantaneous growth rate.
These considerations suggest that the morphology of such kind of AFC Gr-based multilayers might be tailored according to the requirements set by the desired application, which can be relevant for properties such as the coercive field or the exchange coupling \cite{Johnson:RPP:1996,Kuch2006}.

In relation to coercive field, magnetic anisotropy and magnetic remanence engineering, it is well-known that Co/Pd(111), Co/Pt(111) and other related systems such as Co/Ir(111) can display strong PMA with high coercivity and square remanence \cite{Carcia1988a,Johnson:RPP:1996}. 
These properties can be controlled in a considerable extent by the thermal annealing temperature of the intercalation process and the total Co layer thickness.

Thermally activated magnetic hardening in Co/Ir(111) films has been interpreted as a result of the partial interfacial alloying, with maximum coercive fields observed for annealing temperatures around 700$~$K  \cite{Chan2013}, in close proximity to intercalation temperatures here employed. It is thus expected that a certain amount of Co atoms have diffused giving rise to some intermix at the very Co/Ir interface, as recently evidenced by a SXRD investigation of the Co intercalation on partially Gr-covered Ir(111) \cite{Drnec2015}. On the other hand, the preserved integrity of the Graphene layer after Co intercalation (see supp. material S7) avoid Fe/Co intermixing at the top Co or bottom Fe interfaces in room temperature deposition.
Indeed, for a reasonably thin Fe coverage the Co magnetization loop is almost not affected by the Fe deposition, showing the same XMCD at saturation and coercivity value as the bare Gr/Co[1.9ML]/Ir(111). This evidences that the Gr spacing layer effectively preserves chemical and magnetic state of the layers, besides mediating a magnetic coupling through a single atomic spacing layer, which is advantageous for engineering magnetic properties in a superposition scheme. 

\begin{figure*}[t]
\centering
\includegraphics[width=1.0\textwidth]{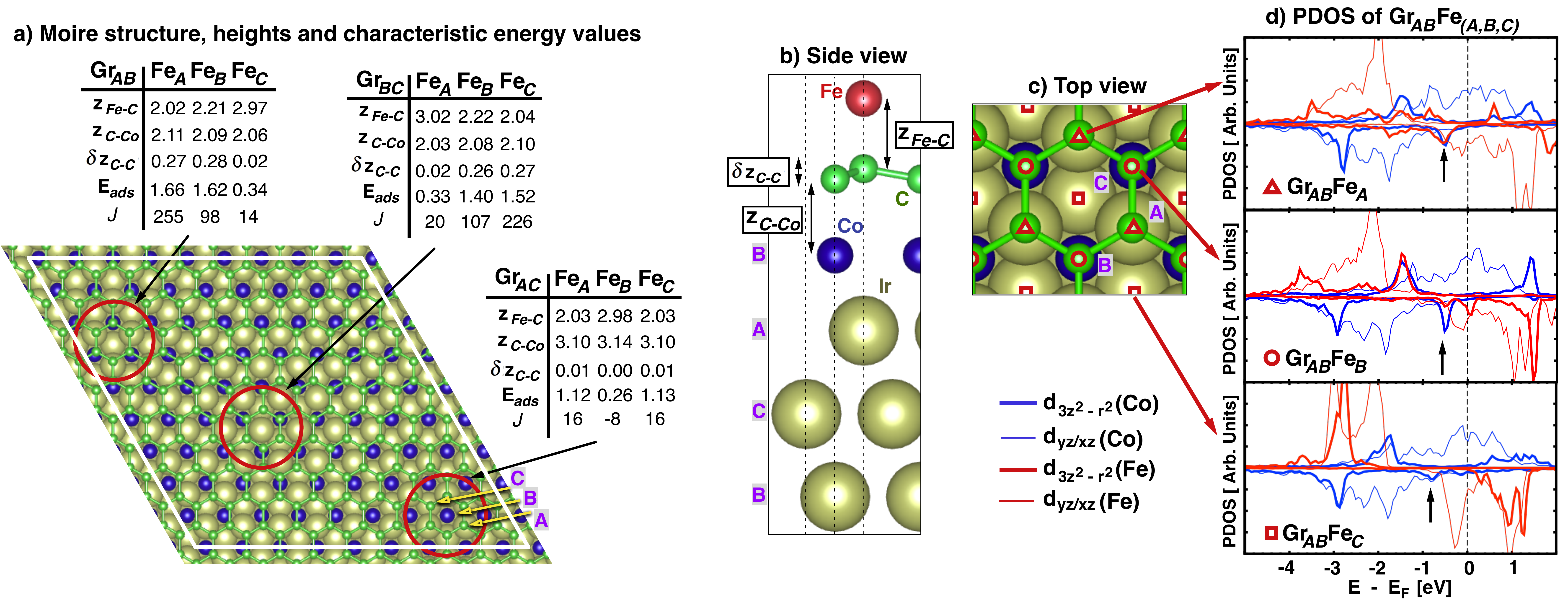}
\caption{\label{fig:moire}\textbf{Structural models for Fe/Gr/Co/Ir(111) heterostructures}. (a) Schematic top view of the Moir\'e superlattice defined by graphene on top of Co/Ir(111). Co is assumed to accommodate in the Ir lattice (in the simulations we neglect intermixing). Three distinct regions can be selected, that we name Gr$_{AB}$, Gr$_{AC}$, and Gr$_{BC}$, and the characteristic interlayer distances, absorption energies and exchange coupling are given (in \AA\ and meV, respectively). (b) It Shows a side view for the Gr$_{AB}$Fe$_A$ configuration, and depicts the stacking adopted throughout the paper for the Ir(111) lattice and for the different MLs' sites~($A,B,C$). Co is always placed at the $B$ site, with terminal Ir on the $A$ site. For the selected commensurate lattice, there are two carbon atoms per Ir/Co/Fe, and we use the notation Gr$_{AB}$ to illustrate that the carbon atoms are placed on $A$ and $B$ sites. Fe monolayer is defined either in $A$, $B$, or $C$ sites, as shown in top view (c) with red triangles, circles and squares, respectively. The pDOS on Co and Fe $d_{3z^2-r^2}$ and $d_{yz/xz}$ for the three posible Fe stacks on top of Gr$_{AB}$ is shown in (d).}
\end{figure*}

To investigate the nature of the coupling in the Graphene-based AFC artificial structure, we performed atomistic calculations based on density functional theory~(DFT), and analysed the role played by the single-layer carbon spacer. 

Our calculations are performed on a model for a Fe/Gr/Co ultra--thin film stacked on a Ir(111) surface, and 
reproduce both the AFC and the strong PMA observed in experiments. 
Due to the lattice mismatch between graphene and Ir(111), a Moir\'e pattern is expected in the heterostructure (the cobalt layer can be considered pseudomorphic with Ir lattice), with a lattice parameter of $\sim$ 2.5$\,$nm (Figure \ref{fig:moire}). Rather than simulating this superstructure (with the 10$\times$10 graphene unit cell and 9$\times$9 Co/Ir lattice, which remains a challenge for atomistic first--principles calculations), we consider a computationally more efficient approach, where commensurability is assumed, and different stacking configurations are used to model the three principal Moir\'e domains sketched in the figure. On top of these, a Fe monolayer is considered, with Fe atoms placed on the possible $A/B/C$ sites for the underlying Ir(111) lattice. Notice that we have labeled the Fe adsorption sites in italic in order to avoid the possibility to misunderstanding the usual C atom symbol. We obtain interlayer distances in good agreement with previous calculations for Co-intercalated in graphene, as reported in panel (a). Notice the strong corrugation in the Moir\'e lattice due to different z$_{\text{C-Co}}$ interlayer distances, of up to 1~\AA. These results on spatial variations for the couplings between magnetic layers, and different PMAs.
The theoretically calculated PMA energies for Gr/Co/Ir(111) interfaces, in the range 2.1--4.5 meV, seem in good agreement with the experimental observations, although they are in contrast to the calculations by Shick et al. \cite{Shick2014} that predict an in--plane magnetization for $AB$ and $BC$ stacks.

In the following, we focus our attention on the Gr$_{AB}$/Co$_B$ stacking, that has a short interlayer distance, and a significant charge accumulation on graphene, which suggests that clustering of Fe adatoms in this region would be preferred during film growth~\cite{Yazyev2010}. The computed Fe--monolayer absorption energies, E$_{ads}$, evaluated after subtracting from the total energy of each configuration the energy of the clean Gr/Co/Ir slab and the Fe ML, conform to this assumption, with Gr$_{AC}$ being less favorable than $AB$ or $BC$.

For the three possible Fe--stacks considered, a strong interlayer exchange coupling, defined as the difference between the energies of parallel (FM) and antiparallel (AF) alignments of magnetizations in Fe and Co ($J$=$E_{FM}-E_{AF}$), of up to a remarkable value of 255$\,$meV, much larger than the values predicted for symmetric Co/Gr/Co or Fe/Gr/Fe junctions~\cite{Yazyev2009}. We mention in passing that similarly strong couplings are obtained for Gr$_{BC}$/Co$_B$ stacks, while the coupling is substantially reduced for Gr$_{AC}$/Co$_B$, where the interlayer distance is larger, and the charge accumulation on graphene suppressed. Computed PMAs show minor differences, hinting that the anisotropy is dominated by intercalated Co.


The origin of the coupling between Fe and Co can be tracked down from the analysis of the electronic structure of the heterointerface. Figure \ref{fig:moire}(d) shows the projected Density of States~(pDOS) on the metallic $d$ states that point out of the layer plane ($d_{3z^2-r^2}$ and $d_{yz/xz}$). It is known that hybridization with graphene´s 2$p_z$ states strongly affects the energy position of Co--3$d$ states~\cite{Yang2016, Friedrich2016}.
A $d_{3z^2-r^2}$ peak at $\sim$0.5$\,$eV below the Fermi level is apparent for both Co and Fe, in all atomic arrangements that give larger couplings, and its amplitude correlates with the value of the $J$ couplings. Notably, if we take the same fixed structures but removing the graphene monolayer, the peaks disappear, revealing that superexchange through C--$p_z$ states is key, and graphene does not act as a mere spacer. Indeed, for the Co--Fe interlayer distances obtained, the coupling becomes FM in absence of graphene, and is reduced by an order of magnitude. Furthermore, calculations done for bilayer and trilayer graphene spacers result in tiny but FM coupling, suggesting that the behaviour of the spacer departs from that of a semimetal, where AF couplings with long decay lengths are obtained~\cite{deVries1997,Pruneda2001}. The strong distortion of graphene´s electronic structure due to the hybridization with the transition metals can be related to this observation. The remarkable buckling $\delta$z$_{\text{C-C}}$ in the graphene monolayer is another consequence of this hybridization.


Having presented both experimental and theoretical evidences establishing the realization of Graphene-mediated AFC FM/Gr structures, we move on to discuss the robustness of this AFC behavior against temperature or layer-thicknesses variations. The stability of the AFC in a broad temperature range is a desirable property in the design of SFI/SAF layered systems.

\begin{figure*}[t]
\includegraphics[width=0.9\textwidth]{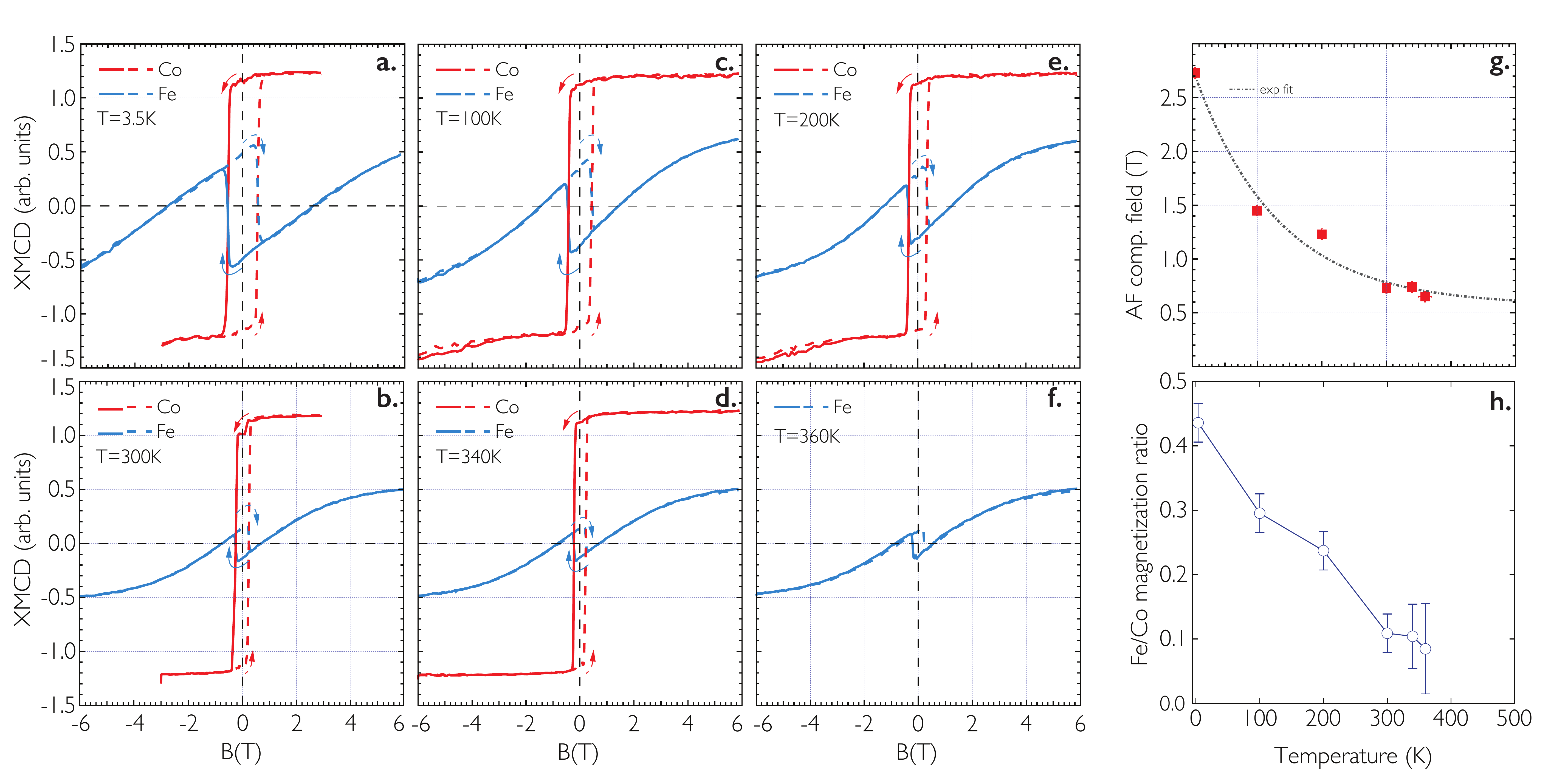}
\centering
\caption{\label{fig:loops_vs_T}\textbf{Temperature dependence of the AFC for a representative Fe[0.9$\,$ML]/Gr/Co[1.9$\,$ML]/Ir(111) sample.} a-f) Element-resolved hysteresis loops collected at increasing sample temperature T=3.5,100,200,300,340,360$\,$K respectively (at T=360 K the Co loop has not been acquired, however the sign inversion of Fe loop still denotes the AFC); g) AF-compensating field, defined as the field B at which the Fe magnetization signal crosses zero, as a function of temperature for the same set of measurement; e) temperature dependence of the Fe/Co magnetization ratio as estimated via the XMCD sum rules.}
\end{figure*}

The set of Fe and Co magnetization hysteresis loops in Figs.\ref{fig:loops_vs_T}(a-e) covering a wide temperature range from T=3.5$\,$K up to T=360$\,$K, reveals a remarkable stability of the AFC state for a Fe[0.6ML]/Gr/Co[1.9ML]/Ir(111) sample grown at room temperature.
As the temperature is lowered, an increase in the out-of-plane Fe layer remanent magnetization occurs, also accompanied by a notable increase in the Co coercive field. The AFC state becomes more robust as the temperature is lowered as evidenced by the enlarged AF-compensating field with decreasing temperatures, that shows an exponential evolution reaching a value of $\sim$3.5T at T=3.5K (see Fig.\ref{fig:loops_vs_T}(f)); however the AF-compensating field keeps a value higher than 0.5$\,$T over the whole temperature range probed and up to 360$\,$K. These temperature trends might result from increases in the magnetic anisotropy, the magnetic susceptibility and magnetic moment of the Fe layer, the exchange coupling strength, or a combination of several of these factors.

In order to asses the capability and design flexibility to realize a compensated Graphene-based SAF/SFI system at a goal temperature (eventually room temperature), we have performed a study of a Fe/Gr/Co/Ir(111) multilayer as a function of its Fe toplayer thickness. We have taken special care towards analyzing two relevant scenarios: i) a magnetically compensated system at room temperature, as the main goal; ii) a structure whose Fe toplayer magnetic moment exceeds that of the Co layer, as inverse scenario of that of an AF structure with a dominant PMI Co layer so far discussed. 

\begin{figure}
\includegraphics[width=\textwidth]{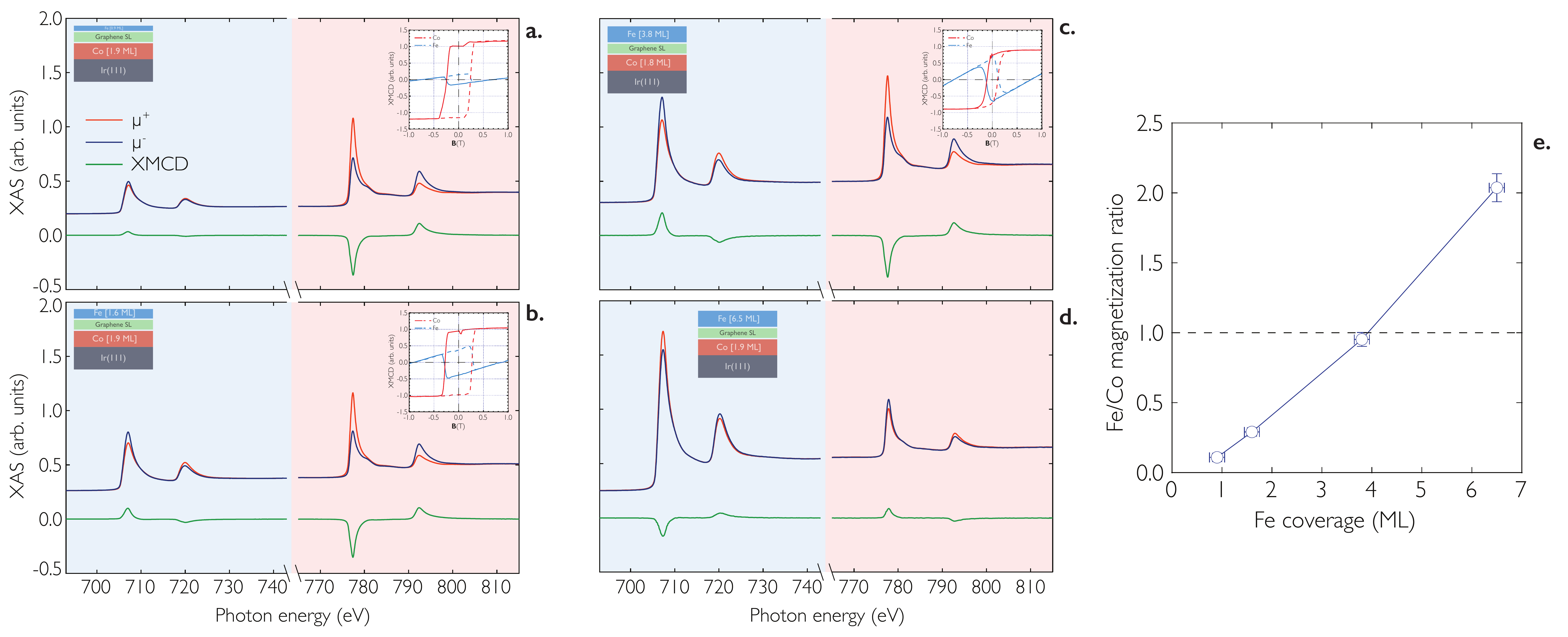}
\centering
\caption{\label{fig:XMCD_vs_cov}\textbf{Element specific XMCD and hysteresis loop (insets) for samples with different Fe coverages at room temperature.} a-d Fe and Co XMCD spectra collected on increasing Fe coverage samples, Fe[0.9 ML]-Co[1.9 ML], Fe[1.6 ML]-Co[1.9 ML] and Fe[3.8 ML]-Co[2.1 ML] and Fe[6.5 ML]-Co[1.9 ML] respectively. The XMCD spectra are collected after applying a field of $\mathbf{B}=+6T$ and subsequently ramping down to 0T in order to fully magnetize the Co layer. In sample reported in panel c) a slightly lower thermal-induced Co intercalation temperature was used, nominally T=640K resulting in a lower coercive field as compared to the other samples. An almost complete magnetic compensation of the perpendicular component is achieved in sample c) as deduced by the XMCD sum rules. The Fe XMCD sign inversion in panel d) indicates that the Fe moment direction is fixed by the external field rather than by the exchange coupling to the Co layer. Nevertheless the AFC is still present as evidenced by the opposite sign of the XMCD between Fe and Co; e) Fe coverage-dependence of Fe/Co magnetization ratio as estimated via the XMCD sum rules (data relative to spectra reported in panels a-d.}
\end{figure}

In Fig.\ref{fig:XMCD_vs_cov}a-d) we report a selected set of Fe and Co $L_{2,3}$ XMCD spectra collected at T=300$\,$K in remanent state for increasing Fe layer nominal coverage on Gr/Co[1.9-2.1 ML]/Ir(111) films. As evidenced by the XMCD sign inversion between Fe and Co atomic edges, the AFC is stable among the different coverages. Most notably, the Fe XMCD signal-to-XAS ratio measured at H=0 increases monotonically up to a 3.8 ML Fe layer thickness, indicating that the remanent Fe magnetization can depends on the Fe layer thickness. Owing to the atomic specificity of the XMCD technique and the difficult accurate determination of the number of atoms probed, it is not straightforward to asses the absolute magnetization of the Fe and Co layers. 
However, we can give a sound estimation of the total Fe/Co out-of-plane magnetization ratios employing the XMCD sum rules \cite{Chen1995} under the plausible assumption that the number of atoms contributing to the XMCD signal is proportional to the Fe and Co coverage. In Fig.\ref{fig:XMCD_vs_cov}(e) we report the room-temperature Fe/Co magnetization ratios as a function of the Fe coverage deduced from spectra reported in panels (a-d). The magnetization ratio show a strong dependence on the Fe coverage demonstrating the total magnetization can be controlled at constant Co layer thickness by tuning the Fe layer coverage. Most notably the Fe/Co magnetization ratio can by tuned to a value close to 1 for a Fe[3.8 ML]/Gr(SL)/Co[2.1 ML]/Ir(111) sample (fig.\ref{fig:XMCD_vs_cov}(c)), suggesting that magnetic compensation can be achieved at room temperature. Following the same analysis based on the XMCD sum rules on the temperature-dependent data set we can get the evolution of the Fe/Co magnetization ratio reported in Fig.\ref{fig:loops_vs_T}h) indicating that the sublayers relative magnetization is strongly influenced by sample temperature. These observations indicate that by sublayer thickness tuning and temperature analysis it is possible to achieve a determined magnetization ratio in a chosen temperature range.

Interestingly, subsequent increase of the Fe sublayer thickness results on an inverted behaviour of Co and Fe sublayer remanent magnetizations (fig.\ref{fig:XMCD_vs_cov}(d)): the Fe out-of-plane remanent magnetization becomes fixed by the direction of the previously applied maximum external field as deduced from the fact that the sign of Fe XMCD is similar for the situations of maximum applied field and zero applied field; in contrast, the Co remanent magnetization now switches and orients antiparallel to the Fe magnetization, presenting an XMCD sign reversal (see fig supp ..). This can be understand by considering that a strong AFC is still in place enforcing the antiparallel alignment of sublayers magnetization at zero field, but that the now much larger Fe toplayer thickness and consequently total magnetization reverse the energy balance of the system towards maintaining unaltered the Fe magnetization direction and switching the Co one. It also reflects that the magnetic balance is changed reversing the roles for the Fe and Co sublayers: now the Fe sublayer drives magnetically the system, whereas the Co layer follows. It is worth noting that we observe an appreciable reduction of the Fe and Co remanent magnetization as deduced from the XMCD intensity in Fig.\ref{fig:XMCD_vs_cov}d), which we argue might be possibly related to the formation of magnetic domains on the Iron layer and via AFC also on the Cobalt layer, resulting in lower sublayer magnetizations in the remanent state.
 
The results for temperature- and thickness-dependency discussed above evidence that, as occurs for a ferrimagnetic material and some conventional SFI systems, it is possible to design Gr-based SFI structures which will present a vanishing net remanent magnetization condition at a defined compensation temperature. A Graphene-based SAF system with compensation of its remanent net magnetization possess the appealing property of having a minimum, ultimately vanishing, stray macroscopic magnetic field. This is a magnetic configuration with maximum stability, as it minimizes its magnetostatic or demagnetization energy. In combination with a high PMA, a large coercive force, the unique Graphene electronic and mechanic properties, and large room-temperature sublayer Fe, Co remanent magnetization, this confers such a Gr-based AFC system properties that appear well-adapted to many of the requirements of magnetic storage media, all-optical magnetic layered materials, or materials for use in magnetic sensors and potential spintronic devices.

In conclusion, we have addressed the feasibility of strong perpendicular AFC in Graphene single-layer spaced ultrahin-film structures. We have experimentally realized Fe/Gr/Co on Ir(111) AFC heterostructures reuniting unprecedent magnetic properties, hence enabling {\it Graphene-based synthetic antiferromagnet and ferrimagnet structures} that appear suitable for applications. Our atomistic calculations confirmed both the AFC and the strong PMA for Fe/Gr/Co structures, revealing that Graphene acts not only as mere spacer but has a direct role in sustaining AF superexchange-coupling between the magnetic layers. These theoretical results provide further ground to the experimental feasibility of Graphene SAF structures, and show that this type of AF coupled structures can be correctly predicted in spite of required approximations and simplifications in their theoretical modeling. This is a remarkable conclusion, as it implies that the development of further structures and engineering of their magnetic properties can be guided or assisted by theory. Altogether, our results demonstrate a novel class of synthetic-antiferromagnetic multilayered materials with remarkable robustness of their AFC at different temperatures and coverage, concomitant with high magnetic remanence and coercive fields. These constitute atomically-thin systems, based on an almost perfect and ultimately thin 2D-layer Graphene spacer with unique electronic and mechanical properties, which might allow a novel range of systems and potential benefits: economy through ultimately thinned down thickness, alternative for rare-earth free structures, novel mechanical properties and possibility for integration on flexible substrates, enhanced interfaces that are sharper and chemically better defined thanks to the protecting effect of the Graphene spacer \cite{Martin2014}, novel interplays between AFC materials and Graphene electron transport properties. Noteworthy, the potential use of such structures in devices would be facilitated by the fact that these systems seem scalable and Silicon integrable, as high-quality magnetic Co, Fe and CVD Gr layers can be grown on Ir(111) films on YSZ buffered Si(111) wafers \cite{Schlenhoff2015}.

The present results open a Graphene-based route for the design of novel synthetic antiferromagnetic materials which, in addition to enclosing fundamental interest, appear of potential use in applications. We expect these results will help spark interest towards the search and discovery of further AFC Gr-based magnetic multilayers with intriguing and remarkable properties, a class of materials largely unexplored and unexploited at present but which could enable new developments in the field of Graphene spintronics.

\section{Methods}

The samples were all prepared in-situ in the preparation chamber available at the HECTOR magnet endstation of the ALBA synchrotron in a pressure better than 5$\times$10$^{-9}mbar$. The Ir(111) single crystal was prepared by repeated cycles of Ar$^+$ ions sputtering at $2KeV$ followed by annealing at T=1000K. The quality of the surface was checked by LEED, giving a sharp six-fold hexagonal pattern without any presence of reconstructions or diffuse background. The Gr/Ir(111) was than prepared exposing the clean substrate held at T=1300K to a C$_2$H$_4$ residual gas atmosphere at a pressure of 2.0$\times$10$^{-6}mbar$ for 10 minutes. The procedure leads to the formation of large single-domain single-layer graphene over the whole surface area, as reported elsewhere and as deduced by Moir\'e LEED pattern (fig. S2) \cite{Coraux2008}. The Gr/Ir(111) samples kept at room-temperature were exposed to a Cobalt flux evaporated from high-purity rod by electron bombardment. The Co deposition rate was about 1\AA/min as determined by a quartz $\mu$-balance. The intercalation process was done either in single step, after having deposited the full amount of Co, or in several intercalation steps at temperatures in the range 570K-700K. Although all the films where showing large PMA and remanent magnetization, larger coercivity values were observed for films prepared either in several steps or at higher annealing temperatures. The full intercalation of Co below the graphene layer was checked by exposing the Gr/Co/Ir(111) to molecular oxygen at a pressure of 1$\times$10$^{-6}mbar$ for 5 min an then checking the XAS at the Co $L_{2,3}$ edge. In case of non-completed intercalation the sample was showing definite signs of Co-oxidation (see supplementary material), while for complete intercalation the sample was shoving a pristine Co $L_{2,3}$ absorption edge. The Fe was deposited at room temperature by electron bombardement evaporation from a high-purity (99.999\%) rod.

The X-ray absorption experiment were carried out at the Boreas beamline of the ALBA synchrotron using a fully circularly polarized X-ray beam produced by an apple-II type undulator\cite{BOREASbeamline:2016}. The base pressure during measurements was $\sim{}1\times10^-10$ mbar. The X-Ray beam was focused to about 500$\times$500$\mu{}m^2$, a gold mesh has been used for incident flux signal normalization. The XAS signal was measured with a Keythley 428 current amplifier as the sample-to-ground drain current (total electron yield TEY signal). The magnetic field was generated collinearly with the incoming x-ray direction by a superconducting vector-cryomagnet (Scientific Magnetics). The magnetization loops where measured sweeping continuosly the magnetic field at a fixed rate and acquiring the absorption TEY current at the maximum of the L$_3$ XMCD signal and at a pre-edge position in order to avoid any field-induced artifact in the measurements.

Our density functional based calculations were performed using the SIESTA code~\cite{Siesta2002}. The generalized gradient approximation~(GGA)~\cite{pbe1996} for the exchange--correlation (XC) potential was considered.
We used norm--conserving pseudopotentials in the separate Kleinman--Bylander~\cite{Kleinman1982} form under the Troullier--Martins parametrization~\cite{Troullier1991}, and to address a better description of the magnetic behavior, nonlinear core corrections were included in the XC terms~\cite{Louie1982}. The geometry optimizations were carried out using the conjugate gradient~(CG) method at spin--polarized scalar relativistic level. A double--$\zeta$ polarized with strictly localized numerical atomic orbitals was used as basis set, and the electronic temperature--kT in the Fermi--Dirac distribution--was set to 5 meV. After the relaxation process the forces per atom were less than 0.01 eV/\AA. The magnetic anisotropy energy~(MAE) were obtained using the on--site Spin--Orbit implementation in SIESTA code~\cite{Seivane2006}. As usual, the MAE is defined as the difference in the total self--consistent energy between hard and easy magnetization directions. Within the present work, we performed an exhaustive analysis of the MAE convergence in order to achieve a tolerance below 10$^{-5}$ eV. We employed around 1000 $k$--points in the calculations for each geometric configuration, which was sufficient to achieve the stated accuracy.

\section{Acknowledgments}
The research leading to this work has been funded by Spanish MINECO/FEDER, grants no. FIS2013-45469-C4-3-R, FIS2015-64886-C5-3-P, Generalitat de Catalunya (2014SGR301), and EU Centre of Excellence “MaX - Materials Design at the Exascale” (H2020 Grant No. 676598). PG, HB and MV acknowledge additional support via ALBA IHR program. MV and PG acknowledge suggestions and feedback from Dr. A. Scholl (LBNL).

\section{Author contributions}

MV and PG conceived the experiments. PG, HB and MV performed sample growth and LEED, Auger, XAS and XMCD experiments at ALBA BL29.  RC and MP designed the theoretical models and performed atomistic calculations. PG and MV analyzed and interpret XMCD data. PG, MV, RC, and MP discussed the magnetic behavior. PG, MV, MP and RC prepared the manuscript.  All authors read and commented on the manuscript.

\bibliography{library}%

\end{document}